# Effect of Coherent State of Quarks and Mesons on Hadron Properties in the Extended Linear Sigma Model

M. Abu-Shady

Institut für Theoretische Physik, Universität Tübingen, D-72076 Tübingen, Germany.

## Abstract
The extended chiral sigma model with quark fields and elementary pion and sigma fields is used to describe static nucleon properties. The field equations have been solved in coherent pair approximation. Better results are obtained for nucleon properties in comparison with previous work and reasonably agree with data.

# Introduction

One of the effective models in describing hadron properties is the linear sigma model which has been suggested earlier by Gell-Mann and Levy [1] to describe nucleons interacting via sigma ($\sigma$) and pion ($\pi$) exchanges. Hadron models set up to understand the structure of nucleon should respect the constraints imposed by chiral symmetry. Spontaneous and explicit chiral symmetry breaking require the existence of the pion whose mass vanishes in the limit of zero current mass. A few solutions for the Lagrangian of chiral linear soliton models when applied to the nucleon and delta already been suggested. Birse and Banerjee [2] solved the linear chiral sigma model in the mean field approximation using the hedgehog ansatz for the pion field. After the variation they performed an approximate projection on angular momentum and isospin ignoring in this procedure the contribution of the pions. Birse [3] and Golli and Rosina [4] have evaluated this model further, performing proper projections even before the variation in the hedgehog approximation. Fiolhais et al. [5] generalized the hedgehog and performed spin and isospin projections as well. Since these authors used exactly the same Lagrangian of Birse and Banerjee [2]. To test the quantum effects, Goeke et al. [6] obtained a static solitonic solution of the linear sigma model using a coherent pair trial Fock state with proper spin and isospin quantum numbers. In contrast with the mean-field approximation the coherent-pair approximation provides a systematic expansion method for the description of a boson field. In addition, it avoids assumptions like the hedgehog structure of the quark and pion fields. The work of Goek et al. [6] have been reexamined by Aly et al. [7] and they corrected some misprints of this work.

In recent years, there has been growing interest in studying nucleon properties, therefore some modifications have been suggested in the linear sigma model in the framework of some aspects of QCD. Broniowski and Golli [8] analyzed a particular extension of the linear sigma model coupled to valence quarks, which contained an additional term with two ingredients of the chiral fields and investigated the dynamic consequences of this term and its relevance to the phenomenology of soliton models of the nucleons. Dmitrasinovic and Myhrer [9] used an extended linear sigma model [10], in which a pair of extra terms has been added to the original linear sigma model in order to improve pion-nucleon scattering and the nucleon sigma term. Also, Korchin [11] calculated the properties of the nucleon in a non-local sigma model, where the conserved electromagnetic and vector currents and the partially conserved axial vector current are obtained. In the same direction, Rashdan et al. [12-14] considered higher-order mesonic interactions in the linear sigma model using mean-field approximation to get a better description of the nucleon properties. Abu-shady [15] suggested mesonic logarithmic potential to calculate nucleon properties in the frame of some aspects of QCD.

The aim of this paper is to estimate the effect of coherent state on hadron properties in the approximating chiral quark sigma model that suggested by Broniowski and Golli [8].

This paper is organized as follows. In Sec.2, the approximating chiral sigma model is explained briefly. The Fock state in the coherent-pair approximation and the variational principle are presented in Secs. 3 and 4, respectively. The derived nucleon properties is explained in Sec. 5. The numerical calculations and discussion of the results are presented in Sec. 6.

# Approximating Chiral Quark Sigma Model

We begin with the approximating chiral quark sigma model [8], which the Lagrangian density of approximating sigma model, which describes the interactions between quarks via the $\sigma$–and $\boldsymbol{\pi}$–mesons is written as in Ref. [8]

$$L(x) = i\overline{\Psi}\partial_\mu\gamma^\mu\hat{\Psi} + \frac{1}{2}(\partial_\mu\hat{\sigma}\partial^\mu\hat{\sigma} + \partial_\mu\hat{\boldsymbol{\pi}}\cdot\partial^\mu\hat{\boldsymbol{\pi}}) + \frac{1}{2}A_0(\sigma\partial^\mu\sigma + \hat{\boldsymbol{\pi}}.\partial^\mu\hat{\boldsymbol{\pi}})^2 + g\,\overline{\hat{\Psi}}\,(\hat{\sigma} + i\gamma_5\hat{\boldsymbol{\tau}}.\hat{\boldsymbol{\pi}})\hat{\Psi} - U(\hat{\sigma},\hat{\boldsymbol{\pi}}) \tag{1}$$

with

$$U(\hat{\sigma},\hat{\boldsymbol{\pi}}) = \frac{\lambda^2}{4}(\hat{\sigma}^2 + \hat{\boldsymbol{\pi}}^2 - v^2)^2 - f_\pi m_\pi^2\hat{\sigma} \tag{2}$$

$$\lambda^2 = \frac{m_\sigma^2 - m_\pi^2}{2f_\pi^2}, \tag{3}$$

$$v^2 = f_\pi^2 - \frac{m_\pi^2}{\lambda^2}, \tag{4}$$

$$\bar{m}_\sigma^2 = (1 + f_\pi^2 A_0)m_\sigma^2 \tag{5}$$

where $f_\pi$ is the pion decay constant, $m_\pi$ is the pion mass, and $m_\sigma$, $g$ and $A_0$ are constants to be determined, wherever $A_0$ is constrained by Eq. 5 ($\bar{m}_\sigma^2 \succ 0$). The quark, sigma and $\boldsymbol{\pi}$- mesons are quantum fields denoted by (^). Spontaneous symmetry breaking generates mass for the quark, which breaks the chiral symmetry and generates the small pion mass which would be zero otherwise as the Goldstone boson of the theory (For details, see Ref. [ 8 ]).

Now, we can write the Hamiltonian density as in Ref. [7].

$$\hat{H}(r) = \frac{1}{2}\{\hat{P}_\sigma(r)^2 + (\nabla\hat{\sigma}(r))^2 + \hat{P}_\pi(r)^2 + (\nabla\pi(r))^2 + A_0(\sigma\partial^\mu\sigma + \hat{\boldsymbol{\pi}}.\partial^\mu\hat{\boldsymbol{\pi}})^2\} +$$

$$U(\hat{\sigma},\hat{\boldsymbol{\pi}}) + \hat{\Psi}^\dagger(r)(-i\alpha\nabla)\,\hat{\Psi}\,(r) - g(r)\hat{\Psi}^\dagger(r)(\beta\hat{\sigma}(r) + i\beta\gamma_5\hat{\boldsymbol{\tau}}.\hat{\boldsymbol{\pi}})\hat{\Psi}(r), \tag{6}$$

where $\alpha$ and $\beta$ are the usual Dirac matrices. In the above expression $\overline{\hat{\Psi}},\hat{\sigma},$ and $\hat{\boldsymbol{\pi}}$ are quantized field operators with the appropriate static angular momentum expansion [7],

$$\hat{\sigma}(r) = \int_0^\infty \frac{d^3k}{[(2\pi)^3 2w_0(k)]^{\frac{1}{2}}}[\hat{c}^\dagger(k)e^{-ik.r} + \hat{c}(k)e^{+ik.r}], \tag{7}$$

$$\hat{\boldsymbol{\pi}}(r) = \left[\frac{2}{\pi}\right]^{\frac{1}{2}}\int_0^\infty dk k^2 \left[\frac{1}{2W_\pi(k)}\right]^{\frac{1}{2}}\sum_{lmw}j_l(kr)Y_{lm}^*(\Omega_r)[\hat{a}_{lm}^{1w\dagger}(k)$$

$$+ (-)^{m+w}\hat{a}_{l-m}^{1-w}(k)], \tag{8}$$

$$\hat{\overline{\Psi}}(r) = \sum_{njmw}\left(\langle r\mid njmw\rangle\hat{d}_{njm}^{\frac{1}{2}w} + \langle r\mid \bar{n}jmw\rangle\hat{d}_{njm}^{\frac{1}{2}w\dagger}\right), \tag{9}$$

where the $|njmw\rangle$ and $|\bar{n}jmw\rangle$ form a complete set of quark and antiquark spinors with angular momentum quantum numbers and spin-isospin quantum numbers $j, m,$ and $w,$ respectively. The corresponding conjugate momentum fields have the expansion [7],

$$\hat{P}_\sigma(r) = i\int_0^\infty d^3k\left[\frac{W_\sigma(k)}{2(2\pi)^3}\right]^{\frac{1}{2}}\left[\hat{c}^\dagger(k)e^{-\mathbf{k}.\mathbf{r}} - \hat{c}(k)e^{+\mathbf{k}.\mathbf{r}}\right],$$

$$\hat{P}_\pi(r) = i\left[\frac{2}{\pi}\right]^{\frac{1}{2}}\int_0^\infty dk k^2\left[\frac{W_\pi(k)}{2}\right]^{\frac{1}{2}}\sum_{lmw}j_l(kr)Y_{lm}^*(\Omega_r)[\hat{a}_{lm}^{1w\dagger}(k) -$$

$$(-)^{m+w}\hat{a}_{l-m}^{1-w}(k)]. \tag{10}$$

Here the $\hat{c}(k)$ destroys a $\sigma$-quantum with momentum **k** and frequency $W_\sigma(k) = (k^2 + m_\sigma^2)^{\frac{1}{2}}$ and $\hat{a}_{lm}^{1w}(k)$ destroys a pion with momentum **k** and corresponding $W_\pi(k) = (k^2 + m_\pi^2)^{\frac{1}{2}}$ in isospin-angular momentum state $\{lm; tw\}$.

## The Fock State

For convenience one constructs the configuration space pion field functions needed for the subsequent variational treatment by defining the alternative basis operators,

$$\hat{b}_{lm}^{1w} = \int dk k^2 \zeta_l(k) \hat{a}_{lm}^{1w}(k), \tag{11}$$

where $\hat{a}_{lm}^{1w}(k)$ are basis operators which create a free massive pion with isospin component $w$ and orbital angular momentum $(l, m)$, and $\zeta_l(k)$ is the variational function. Taking this over to configuration space defines the pion field function [7]

$$\Phi_l = \frac{1}{2\pi} \int_0^\infty dk k^2 \frac{\zeta_l(k)}{W_\pi(k)^{\frac{1}{2}}} j_l(r) \tag{12}$$

In the following only the $l = 1$ value is used and the angular momentum label will be dropped. The Fock state for the nucleon is taken to be [7]

$$|NT_3 J_z\rangle = [\alpha(|n\rangle \otimes |P^{00}\rangle)_{T_3 J_z} + \beta(|n\rangle \otimes |P^{11}\rangle)_{T_3 J_z} +$$
$$\gamma(\delta > \otimes |P^{11}\rangle_{T_3 J_z})|0\rangle]\Big|\sum\Big\rangle, \tag{13}$$

where $|\sum\rangle$ is the coherent sigma field state with the property: $\langle\sum|\hat{\sigma}(r)|\sum\rangle = \hat{\sigma}(r)$, and $|P^{00}\rangle(|P_{1m}^{1w}\rangle)$ are pion coherent-pair states to be determined. The normalization of the nucleon state requires $\alpha^2 + \beta^2 + \gamma^2 = 1$. The permutation symmetric form of that $SU(2) \times SU(2) \times SU(2)$ quark wave functions imply that the source terms in the pion field equations will induce in angular momentum isospin correlation for the pion field. (For details, see Ref. [7])

## The Variational Principle

The objective of this section is to seek the minimum of the total energy of baryon is given by

$$E_B = \langle BT_3 J_z | \int_0^\infty d^3 r : H(r) : |BT_3 J_z\rangle, \tag{14}$$

where $B = N$ or $\Delta$. The field equations are obtained by minimizing the total energy of the baryon with respect to variation of the fields, $\{u(r), w(r), \sigma(r), \Phi(r)\}$, as well as the Fock-space parameters, $\{\alpha, \beta, \gamma\}$ subject to the normalization conditions. The total energy of the system is written as

$$E_B = 4\pi \int_0^\infty dr r^2 \varepsilon_B(r). \tag{15}$$

Writing the quark Dirac spinor as

$$\Psi_{\frac{1}{2}m}^{\frac{1}{2}w}(\mathbf{r}) = \begin{pmatrix} u(r) \\ v(r)\sigma.\hat{\mathbf{r}} \end{pmatrix} \chi_{\frac{1}{2}m} \zeta^{\frac{1}{2}w}, \tag{16}$$

the energy density is given by

$$\varepsilon_B(r) = \frac{1}{2}\left(\frac{d\sigma}{dr}\right)^2 + \frac{1}{2}A_0\sigma^2\left(\frac{d\sigma}{dr}\right)^2 + A_0\Phi\frac{d\Phi}{dr}\sigma\frac{d\sigma}{dr} - \frac{\lambda^2}{4}(\sigma^2(r)-v^2)^2 - m_\pi^2 f_\pi \sigma(r)$$
$$+ 3\left[u(r)\left(\frac{dv}{dr} + \frac{2}{r}v(r)\right) - v(r)\frac{du}{dr} + g\sigma(r)(u^2(r) - v^2(r))\right]$$
$$+ (N_\pi + x)\left(\left(\frac{d\Phi}{dr}\right)^2 + \frac{2}{r^2}\Phi^2(r)\right) + (N_\pi - x)\Phi_p^2(r) -$$
$$\alpha\delta g(a+b)u(r)v(r)\Phi(r) + \lambda^2[x^2 + 2xN_\pi + 81(\alpha^2 a^2 c^2 + (\beta^2+\gamma^2)d^2)]\Phi^4(r) +$$
$$\lambda^2(N_\pi + x)(\sigma^2(r) - v^2)\Phi^2(r) + A_0\Phi(r)^2(N_\pi+x)\left((\frac{d\Phi}{dr})^2 + \frac{2}{r^2}\Phi^2\right) \qquad (17)$$

where $N_\pi$ is the average pion number
$$N_\pi = 9(\alpha^2 a^2 + (\beta^2+\gamma^2)c^2), \qquad (18)$$
and where $\delta$ takes the following values for nucleon or delta quantum numbers,
$$\delta_N = \left(5\beta + 4\sqrt{2}\gamma\right)/\sqrt{3}, \delta_\Delta = \left(2\sqrt{2}\beta + 5\gamma\right)/\sqrt{3} \qquad (19)$$
The function $\Phi_p(r)$ is obtained from $\Phi(r)$ by the double folding,
$$\Phi_p(r) = \int_0^\infty w(r,\acute{r})\Phi(\mathbf{r})r^2 d\acute{r}, \qquad (20)$$
$$w(r,\acute{r}) = \frac{2}{\pi}\int_0^\infty dk k^2 w(k) j_1(kr) j_1(kr'). \qquad (21)$$

For fixed $\alpha, \beta$ and $\gamma$ the stationary functional variations are expressed by
$$\delta\left[\int_0^\infty dr r^2 (\varepsilon_B(r) - 3\epsilon(u^2(r)+v^2(r)) - 2k\Phi\Phi_p(r))\right] = 0, \qquad (22)$$
where the parameter k enforces the pion normalization condition,
$$8\pi \int_0^\infty \Phi(r)\Phi_p(r)r^2 dr = 1, \qquad (23)$$
and $\epsilon$ fixes, the quark normalization,
$$4\pi \int_0^\infty (u^2(r) + v^2(r))r^2 dr = 1. \qquad (24)$$

Minimizing the Hamiltonian yields the four nonlinear coupled differential equations,
$$\frac{du}{dr} = -2(g\sigma + \epsilon)v(r) - \frac{1}{3}\alpha\delta(a+b)g\Phi(r)u(r), \qquad (25)$$
$$\frac{dv}{dr} = -\frac{2}{r}v(r) - 2(g\sigma(r) - \epsilon)u(r) + \frac{1}{3}\alpha\delta(a+b)g\Phi(r)u(r), \qquad (26)$$
$$\frac{d^2\sigma}{dr^2} = \frac{1}{(1+A_0\sigma^2)}\{-\frac{2}{r}(1+A_0\sigma^2)\frac{d\sigma}{dr} - m_\pi^2 f_\pi + 3g(u^2(r) - v^2(r)) + 2\lambda^2(N_\pi + x)\Phi^2(r)\sigma(r)$$
$$+ \lambda^2(\sigma^2(r) - v^2)\sigma(r)\}, \qquad (27)$$
$$\frac{d^2\Phi}{dr^2} = -\frac{2}{r}\frac{d\Phi}{dr} + \frac{2}{r^2}\left(\frac{2N_\pi + A_0\Phi^2(N_\pi+x)}{4N_\pi + 2A_0\Phi^2(N_\pi+x)}\right)\Phi(r) + \left(\frac{2N_\pi + 2x}{4N_\pi + 2A_0\Phi^2(N_\pi+x)}\right)m_\pi^2\Phi(r)$$
$$+ \frac{1}{4N_\pi + 2A_0\Phi^2(N_\pi+x)}\{4\lambda^2[x^2 + 2N_\pi x + 81(\alpha^2 a^2 c^2 + (\beta^2+\gamma^2)d^2)]\Phi^3(r)$$
$$- \alpha(a+b)g\delta u(r)v(r) + 2\lambda^2(N_\pi+x)(\sigma^2 - v^2)\Phi -$$
$$2K\Phi(r)\} + \frac{2A_0\Phi(N_\pi+x)}{4N_\pi + 2A_0\Phi^2(N_\pi+x)}\left((\frac{d\Phi}{dr})^2 + \frac{2}{r^2}\Phi^2\right), \qquad (28)$$

with eigenvalue $\epsilon$ and $k$. These consist of two quark equations for $u$ and $v$ where $\sigma(r)$ and $\Phi(r)$ appear as potentials, and two Klein-Gordon equations with $u(r)v(r)$ and $(u^2(r) - v^2(r))$ as source terms. The boundary conditions are for $r \to 0$,

$$v = \frac{d\sigma}{dr} = \Phi = \frac{du}{dr} = 0, \tag{29}$$

and for $r \to \infty$,

$$\left[ r(gf_\pi - \epsilon)^{\frac{1}{2}} + (gf_\pi + \varepsilon)^{-\frac{1}{2}} \right] u(r) - r(gf_\pi + \epsilon)^{\frac{1}{2}} v(r) = 0, \tag{30}$$

$$(2 + 2m_\pi r + m_\pi^2 r^2)\Phi(r) + r(1 + m_\pi r)\Phi(r) = 0, \tag{31}$$

$$r\sigma(r) + (\sigma(r) - f_\pi)(1 + m_\sigma r) = 0, \tag{32}$$

The field equations are solved for fixed coherence parameter, $x$, and fixed Fock-space parameter,$(\alpha, \beta, \gamma)$ as Ref. [7].

## The Nucleon Properties

The expectation value of the energy is minmized with respect to $(\alpha, \beta, \gamma)$ by diagonalization of the energy matrix

$$\begin{bmatrix} H_{\alpha\alpha} & H_{\alpha\beta} & H_{\alpha\gamma} \\ H_{\alpha\beta} & H_{\beta\beta} & H_{\beta\gamma} \\ H_{\alpha\beta} & H_{\beta\gamma} & H_{\gamma\gamma} \end{bmatrix} \begin{bmatrix} \alpha \\ \beta \\ \gamma \end{bmatrix} = E \begin{bmatrix} \alpha \\ \beta \\ \gamma \end{bmatrix} \tag{33}$$

Each H entry of the matrix is related to a corresponding density, as follows:

$$H_{\alpha\beta} = 4\pi \int r^2 E_{\alpha\beta}(r) dr, \tag{34}$$

and analogously for the other entries. The functions for a nucleon are

$$E_{\alpha\alpha} = E_0(r) + 18a^2 \Phi_p^2(r) + 9a^2 \lambda^2 (2x + 9c^2) \Phi^4(r) + 9a^2 \lambda^2 (\sigma^2(r) - v^2) \Phi^2(r) +$$
$$9A_0 a^2 \Phi^2(r)((\frac{d\Phi}{dr})^2 + \frac{2}{r^2} \Phi^2) \tag{35}$$

$$E_{\beta\beta} = E_0(r) + 18c^2 \Phi_p^2(r) + 9\lambda^2 (2xc^2 + 9d^2) \Phi^4(r) + 9c^2 \lambda^2 (\sigma^2(r) - v^2) \Phi^2(r) +$$
$$9A_0 c^2 \Phi^2(r)((\frac{d\Phi}{dr})^2 + \frac{2}{r^2} \Phi^2) \tag{36}$$

$E_{\alpha\beta}$ and $E_{\alpha\gamma}$ as in Ref. [7]. The equations of Ref. [7] are recovered at $A = 0.0$.
where

$$E_0(r) = \frac{1}{2} \left(\frac{d\sigma}{dr}\right)^2 + \frac{1}{2} A_0 \sigma^2 \left(\frac{d\sigma}{dr}\right)^2 + A_0 \Phi \frac{d\Phi}{dr} \sigma \frac{d\sigma}{dr} - \frac{\lambda^2}{4} (\sigma^2(r) - f_\pi^2)^2$$
$$+ 3g\sigma(r)(u^2(r) - v^2(r)) + (N_\pi + x)((\frac{d\Phi}{dr})^2 + \frac{2}{r^2} \Phi^2(r)) +$$
$$(N_\pi - x)\Phi_p^2(r) - 2xm_\pi^2 \Phi(r)^2 + \lambda^2 x^2 \Phi^4(r) + \frac{m_\pi^2}{4} (\sigma(r) - f_\pi)^2$$
$$\lambda^2 x(\sigma^2(r) - f_\pi^2)\Phi^2(r) + A_0 \Phi(r)^2 (N_\pi + x)((\frac{d\Phi}{dr})^2 + \frac{2}{r^2} \Phi^2) \tag{37}$$

We calculate the properties of the nucleon such as the magnetic moments, nucleon mass and charge radius in fm. The equations unchanged as in Ref. [7] this due to the change in this quantities are induced through the change of the dynamics of fields equations [8].

## Numerical Results and Discussion

The set of nonlinear differential equations have been solved in the same manner as Aly et al. [7]. The iteration procedure is implemented as follows. For fixed values $x, A_0, \alpha, \beta$ and $\gamma$ however, the set of differential equations with the corresponding boundary conditions are solved by using the modified numerical package (COLSYS) as used in Ref. [7]. The solutions of the system are mixed and repeated until self-consistency is achieved.

Sigma field is play important role as partner with pion field in chiral symmetry therefore a few authors interested recently to calculate the sigma mass as Leutwyler [17] in the framework of chiral perturbation theory so we calculate all observables of the nucleon for $m_\sigma = 441$ MeV.

To estimate the effect of the A-term on nucleon properties in coherent pair approximation. We see from Fig. 1 the A-term is lowered the energy of nucleon and delta masses in comparison with original model [7]. This result is agreement with the result in mean-field approximation [8] therefore the stability is achieved at high values of $g$. This is a desired effect since many phenomenological approaches have problems in getting the mass in the right ball park if $g$ is too high [8].

To estimate the effect of coherent state, the degree of the coherent parameter $x$ is taken =3.0 leads to improving in the nucleon properties in comparison with previous work [7](see, Table 2); the charge radius of proton is excellent agreement with measured data and charge radius of neutron is improved in comparison Refs. [7, 16], where the relative error in the neutron radius about 97%. The nucleon magnetic moments are improved in comparison with Ref. [7, 16] in range (10-20)%. Fig. 2. shows the mesons fields for the $x = 3.0$, $g = 5$ and $m_\sigma = 441$ MeV, where the presence of the A-term weakens the pion mass, modifies the shape of the sigma field, as well as slightly increases the size of the soliton that leads to the stability in energy of the soliton therefore the behavior is agreement with the behavior in mean-field approximation as in Ref.[8].

**Table (1)**: The energy contributions (in MeV) to nucleon and delta when using $g = 5$, $m_\sigma = 441$ MeV, $A_0 = 0.9$ MeV$^{-2}$. The coherent parameter is taken $x = 3$ for the present work and the original model [7]

| Quantity | Nucleon | Delta | Nucleon[7] | Delta[7] |
|---|---|---|---|---|
| Quark kinetic energy | 932.246 | 823.16 | 889.38 | 784.31 |
| Sigma kinetic energy | 216.414 | 163.24 | 312.425 | 271.92 |
| Pion kinetic energy | 370.865 | 293.10 | 385.91 | 318.24 |
| Quark-meson interaction | -565.077 | -222.068 | -557.20 | -245.035 |
| Meson interaction energy | 165.91 | 188.71 | 180.829 | 205.62 |
| Baryon mass | 1120.36 | 1246.142 | 1211.34 | 1335.06 |
| Nucleon-Delta mass difference | | 125.784 | | 123.724 |

**Table (2)**: Nucleon observables using $x = 3$, $m_\sigma = 441$ MeV, $A0 = 0.9$ MeV$^{-2}$ and $g = 5$

| Quantity | Quark | Meson | Total | [7] | [16] | Expt. |
|---|---|---|---|---|---|---|
| $r_c^2$(proton)(fm$^2$) | 0.684 | 0.046 | 0.73 | 0.556 | 0.69 | 0.7 |
| $r_c^2$(neutron)(fm$^2$) | 0.028 | -0.108 | -0.08 | -0.004 | -0.03 | -0.12 |
| Magnetic moment (proton) | 1.66 | 0.44 | 2.1 | 1.71 | 1.93 | 2.79 |
| Magnetic moment (neutron) | -1.24 | -0.46 | -1.7 | -1.31 | -1.6 | -1.91 |

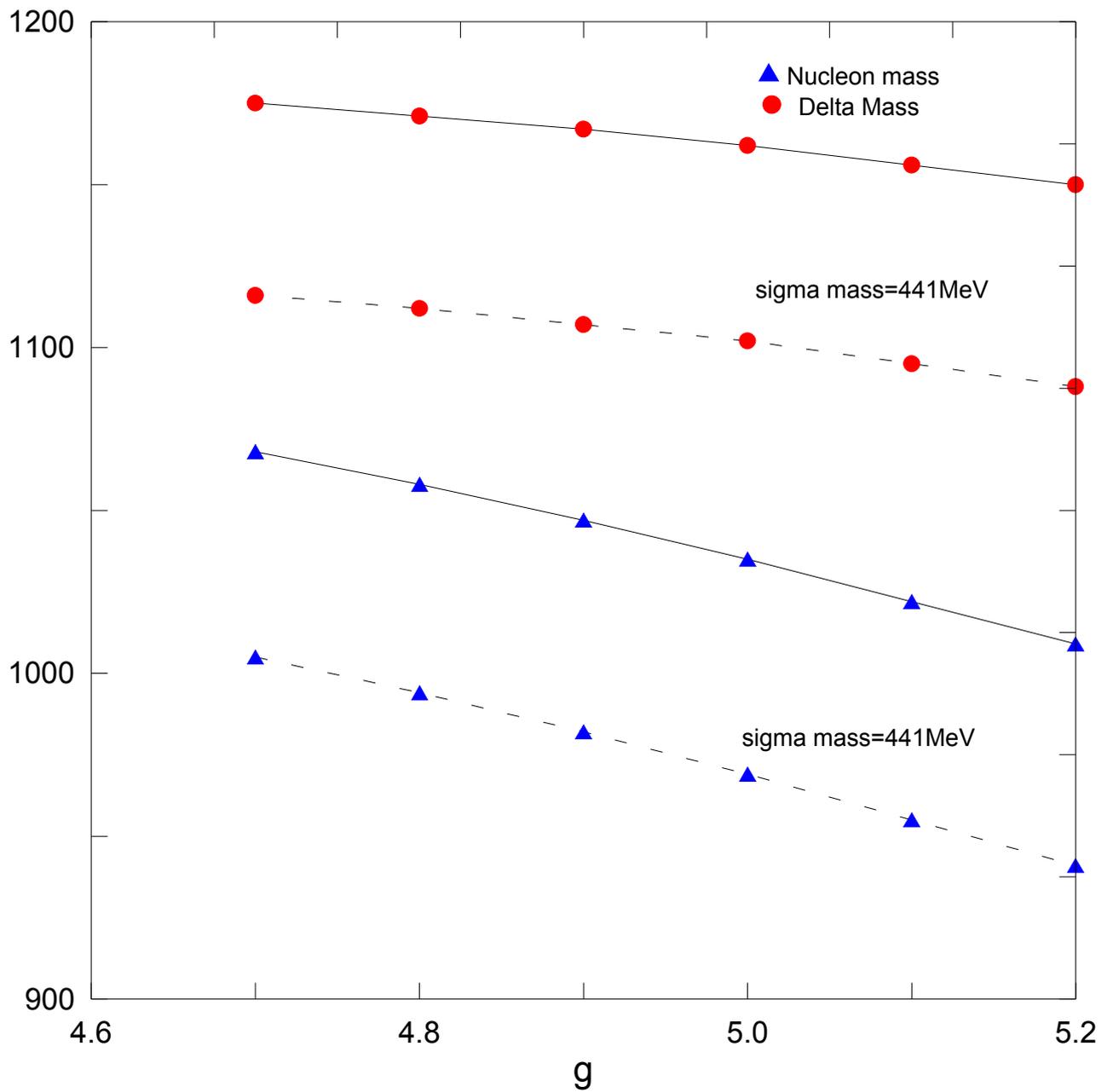

Fig. 1. The dependence of the energy of the soltion on $g$ for $A_0 = 0$ (solid lines) and $A_0 = 0.9$ MeV$^{-2}$ (dashed lines)

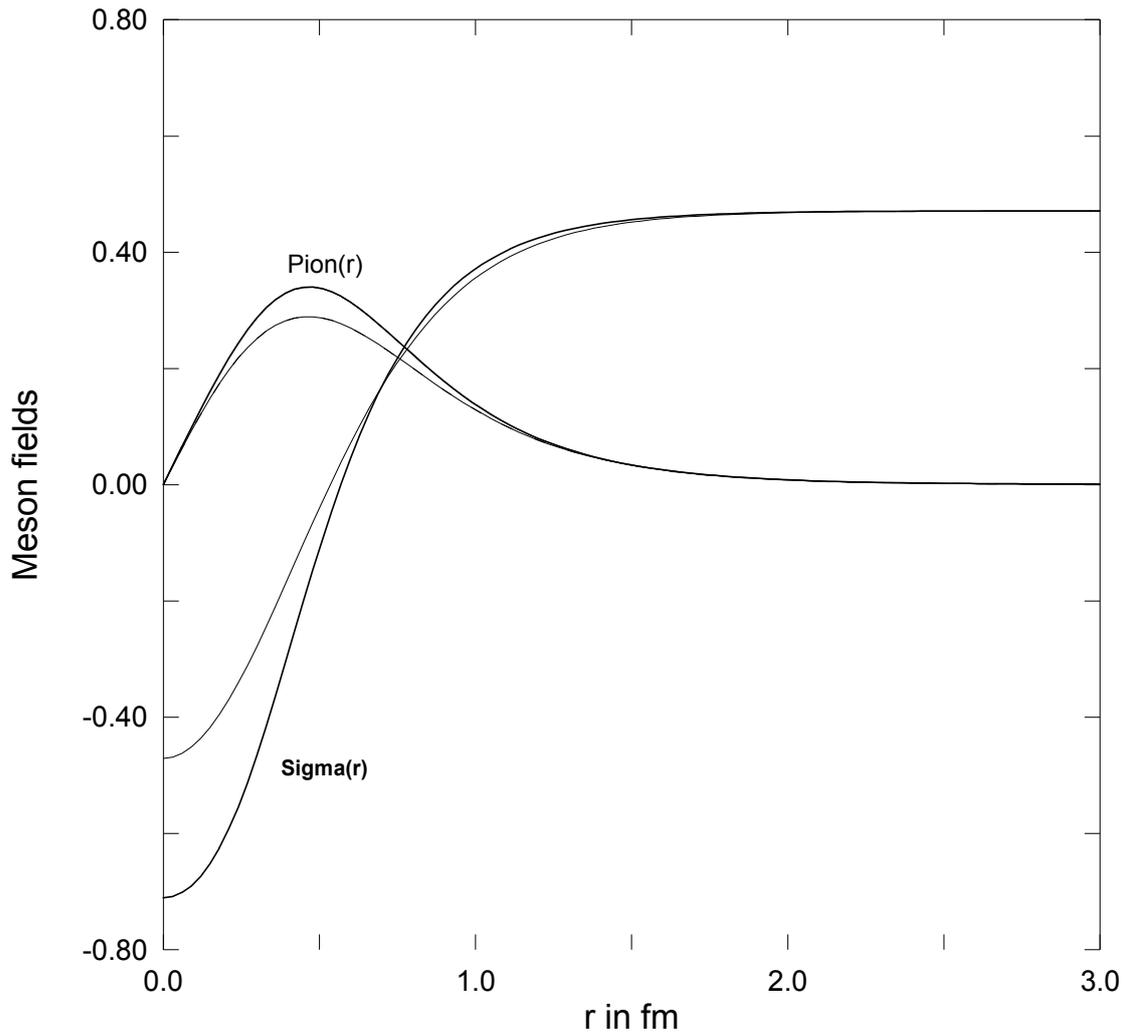

Fig. 2. $\sigma$ and $\pi$ fields for $A_0 = 0$ (bold curves) and $A_0 = 0.9$ MeV$^{-2}$ (normal curves).

# Acknowledgments

This work is financed by the government of Egypt. I would like to thank Prof. A. Faessler for the hospitality during at my stay at Tubingen University and thank Prof. T.S.T. Aly for his comments that supports this work.